\documentclass[english]{article}
\usepackage[T1]{fontenc}
\usepackage[latin1]{inputenc}
\usepackage{geometry}
\usepackage{float}
\usepackage{graphicx}

\makeatletter

\providecommand{\tabularnewline}{\\}

\usepackage{babel}
\makeatother
\begin{document}

\title{Higher-twist analysis of moments of spin structure function}

\author{A. Deur \\
Thomas Jefferson National Accelerator Facility, Newport News, VA 23606}

\maketitle
\begin{abstract}
Available analyses on moments of the spin structure function $g_{1}$
use different methods and are barely consistent with each other. We
present a higher twist analysis of $\Gamma_{1}^{p}$ using a method
consistent with the studies of $\Gamma_{1}^{n}$ and $\Gamma_{1}^{p-n}$
already published. The twist-4 coefficient $f_{2}$ is extracted.
One result is that the higher twist coefficients seem to alternate
signs: the relatively larger twist-6 contribution is partly suppressed
by the twist-4 and twist-8 contributions. The size of twist-6 can
be due to the elastic contribution to the moments. 
\end{abstract}
High precision data on doubly polarized electron-nucleon scattering
from Jefferson Lab (JLab) have been analyzed in the transition regime
from asymptotically free to strongly interacting quarks {[}\ref{EG1a proton},\ref{EG1a Deuteron},\ref{E94010},\ref{E94010-2}{]}.
Studying quark-gluon and quark-quark interactions is important to
understanding quark confinement. Such study can be cast in the Operator
Product Expansion (OPE) formalism, which describes in particular the
evolution of structure functions and their moments. The Cornwall-Norton
moment is the integral of the structure function over $x$. Here,
$x=Q^{2}/2M\nu$ is the Bjorken variable, $Q^{2}$ is the four-momentum
transfer from the electron to the nucleon, $\nu$ is the energy transfer
and $M$ is the nucleon mass. In OPE, the first moment of $g_{1}(x,Q^{2})$
can be written as:

~

$\Gamma_{1}(Q^{2})\equiv\int_{0}^{1}dxg_{1}(x,Q^{2})=\sum_{\tau=2,4...}\frac{\mu_{\tau}(Q^{2})}{Q^{\tau-2}}$,

~

\noindent where the $\mu_{\tau}(Q^{2})$ are sums of twist elements
added up to twist $\tau$. The twist is defined as the mass dimension
minus the spin of an operator. Twist elements$\geq3$ can be related
to quark-quark and quark-gluon interactions. Hence they are important
quantities for confinement study. The leading twist coefficient is:

~

$\mu_{2}^{p(n)}(Q^{2})=C_{ns}(Q)^{2}\left(\frac{1}{36}a_{8}\pm\frac{1}{12}g_{A}\right)+C_{s}(Q)^{2}\frac{1}{9}a_{0}$

~

\noindent where $C_{ns}$ and $C_{s}$ are flavor non-singlet and
singlet Wilson coefficients that represent the $Q^{2}$-dependence
due to QCD radiations {[}\ref{Wilson Coef.}{]}, $g_{a}=1.267(35)$
is the triplet axial charge {[}\ref{axial charges}{]}, $a_{8}=0.579(25)$
is the octet axial charge {[}\ref{axial charges}{]} and $a_{0}$
is the singlet axial charge. In the $\overline{MS}$ renormalization
scheme that will be used here, $a_{0}=\Delta\Sigma$ where $\Delta\Sigma$
is the contribution of the quarks to the nucleon spin. The next to
leading order twist coefficient is:

~

$\mu_{4}(Q^{2})=\frac{M^{2}}{9}\left(a_{2}(Q2)+4d_{2}(Q^{2})+4f_{2}(Q^{2})\right)$

~

\noindent $a_{2}$ ($d_{2}$) is a twist two (three) target mass
correction that can be related to higher moments of $g_{1}$ (of $g_{1}$
and $g_{2}$), and $f_{2}$ is the twist four contribution {[}\ref{Twist-4 SV}{]}. 

OPE analysis can also be carried out using Nachtmann moments {[}\ref{Nachtmann}{]},
in which the target mass corrections are done by an appropriate combination
of $g_{1}$ and $g_{2}$ in the moment's definition. Such an analysis
of the new JLab EG1a data have been carried out on $\Gamma_{1}^{p}$
{[}\ref{Osipenko}{]}. On the other hand, analysis of $\Gamma_{1}^{n}$
{[}\ref{Meziani}{]} and the flavor non-singlet $\Gamma_{1}^{p-n}$
{[}\ref{Bjorken-HT}{]} were done using Cornwall-Norton moments. The
results at $Q^{2}=1$ GeV$^{2}$ for $f_{2}$ are $f_{2}^{p}=0.039\pm0.022$(stat)$_{-0.018}^{+0.000}$(syst)$\pm0.030$(low
$x$)$_{-0.011}^{+0.007}$($\alpha_{s}$), $f_{2}^{n}=0.034\pm0.005\pm0.043$
and $f_{2}^{p-n}=-0.13\pm0.15$(uncor.)$_{-0.03}^{+0.04}$(cor.) where
uncor. (cor.) specifies the error due to the uncorrelated (correlated)
experimental uncertainty. The $\mu_{6}$ results are $\mu_{6}^{p}*/M^{4}=$$0.011\pm0.013$(stat)$_{-0.000}^{+0.010}$(syst)$\pm0.011$(low
$x$)$\pm0.000$($\alpha_{s}$), $\mu_{6}^{n}/M^{4}=-0.019\pm0.002\pm0.017$
and $\mu_{6}^{p-n}/M^{4}=0.09\pm0.06$(uncor.)$\pm0.01$(cor.) where
the asterisk in $\mu_{6}^{p}*$ recalls that this coefficient contains
only a twist 6 term.

These results, while coming from the same set of data, barely agree.
The disagreement could come from the fact that the low-$x$ extrapolation
procedures differ in the three analyses, or the lower $Q^{2}$ considered
for the fits are different (1 GeV$^{2}$ for p, 0.5 for n and 0.8
for p-n), or the target mass corrections are treated differently in
the Nachtmann and Cornwall-Norton analyses: in the Nachtmann moments,
target mass corrections are added to all orders while in the Cornwall-Norton
analyses, only the first order is corrected for. The Cornwall-Norton
analyses indicate that twist 4 and twist 6 terms are of similar magnitude
(although twist 6 is larger) but opposite sign, leading to a partial
cancellation of higher twist effects. This is not as clear from the
Nachtmann analysis. To clarify this issue, it would be beneficial
to provide consistent OPE analysis of the data. In that light, we
have redone a Cornwall-Norton analysis of the $\Gamma_{1}^{p}$ data
consistent with the $\Gamma_{1}^{n}$ and $\Gamma_{1}^{p-n}$ analyses.

The low$-x$ extrapolation of the JLab and world data was redone,
as in the $\Gamma_{1}^{n}$ and $\Gamma_{1}^{p-n}$ analyses, using
the Thomas-Bianchi parametrization {[}\ref{BT}{]} up to the invariant
mass squared $W^{2}=1000$ GeV$^{2}$. The uncertainty was estimated
by varying all the parameters within their range given in {[}\ref{BT}{]}.
A Regge form {[}\ref{regge}{]} was used beyond $W^{2}=1000$ GeV$^{2}$
on which an uncertainty of 100\% was assumed. The elastic contribution
to the moments was estimated using the parametrization of Mergell
\emph{et al}. {[}\ref{FF}{]}. A 2\% uncertainty was assumed. The
JLab EG1a experiment (that will mainly determine the higher twist
magnitude) is dominated by systematic uncertainties. Its point to
point uncorrelated systematic uncertainties were separated from its
correlated ones, and added in quadrature to its statistical uncertainty.
This error was used in the OPE fit. The effect of the point to point
correlated uncertainty was accounted for by shifting the EG1a data
set and using it as a new input in the fit. 

Fitting the world data for $Q^{2}\geq5$ GeV$^{2}$ and assuming no
higher twist effects above $Q^{2}=5$ GeV$^{2}$ yields $\Delta\Sigma=0.154\pm0.066$.
The target mass correction $a_{0}(Q^{2})=\int_{0}^{1}dx\left(x^{2}g_{1}(x,Q^{2})\right)$,
where $g_{1}(x,Q^{2})$ contains only a twist-2 contribution, was
estimated with the parton distribution parameterization of J. Bluemlein
and H. Boettcher {[}\ref{BB}{]}. The twist-3 contribution $d_{2}(Q^{2})$
was obtain from the SLAC E155x experiment {[}\ref{E155x}{]}. Although
accounting for the $Q^{2}$ dependence had little effect on the fit,
a $Q^{2}-$dependence of the form $A(Q^{2})=A(Q_{0}^{2})\left(\alpha_{s}(Q_{0}^{2})/\alpha_{s}(Q^{2})\right)^{b}$
was assumed for $a_{0}(Q^{2})$ and $d_{2}(Q^{2})$ with $b=-0.2$
and $b=-1$ respectively. $\Lambda_{QCD}=0.37_{-0.07}^{+0.04}$ was
used in computing $\alpha_{s}(Q^{2})$.

The world data together with the OPE leading twist evolution (LT)
of $\Gamma_{1}^{p}(Q^{2})$ and the elastic contribution to $\Gamma_{1}^{p}(Q^{2})$
are shown in the figure below. The band at zero is the point to point
correlated uncertainty on the JLab EG1a data. The dot-dashed line
is the result of fit 1 (see table).

\begin{figure}[H]
\begin{center}\includegraphics[%
  scale=0.4]{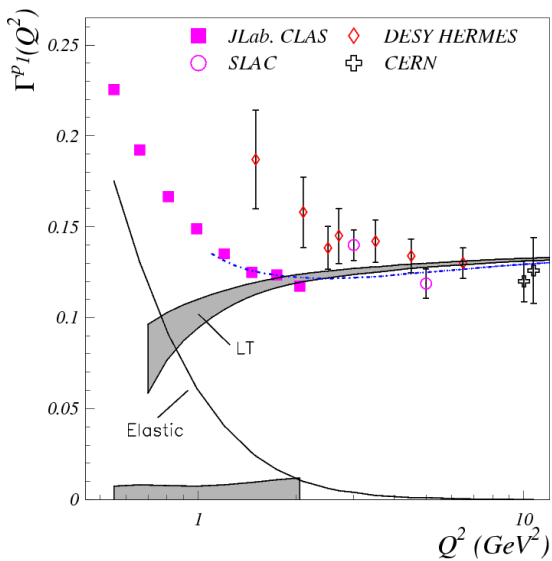}\end{center}

\caption{World data on $\Gamma_{1}^{p}(Q^{2})$. The gray band (LT) is the
pQCD leading twist evolution. The band on the horizontal axis is the
point to point correlated uncertainty for the JLab CLAS experiment.
The uncorrelated uncertainty is of the size of the square symbols.
The error bars on the open symbols are systematic and statistic added
in quadrature. The dash-dotted line is a fit of the data starting
at $Q_{min}^{2}=0.1$ GeV$^{2}$.}
\end{figure}

To check the convergence of the OPE series, the lowest $Q^{2}$ value,
$Q_{min}^{2}$was varied, as well as OPE series truncated to twist-8.
The results are given in the table below. All the twist coefficient
values are given for $Q^{2}$=1 GeV$^{2}$. The first error represents
the uncorrelated uncertainty, coming mainly from the statistical uncertainty,
and the second is the point to point correlated uncertainty.

~

\begin{tabular}{|c|c|c|c|c|c|}
\hline 
{\footnotesize fit}&
{\footnotesize $Q_{min}^{2}$}&
{\footnotesize $f_{2}$ }&
{\footnotesize $\mu_{4}/M^{2}$}&
{\footnotesize $\mu_{6}/M^{4}$}&
{\footnotesize $\mu_{8}/M^{6}$}\tabularnewline
\hline
\hline 
{\footnotesize 1}&
{\footnotesize 1.0}&
{\footnotesize -0.138$\pm$0.024$_{-0.101}^{+0.113}$}&
{\footnotesize -0.055$\pm$0.011$_{-0.046}^{+0.050}$}&
{\footnotesize 0.110$\pm0.014$$_{-0.046}^{+0.041}$}&
{\footnotesize -}\tabularnewline
\hline 
{\footnotesize 2}&
{\footnotesize 0.8}&
{\footnotesize -0.120$\pm0.017{}_{-0.015}^{+0.091}$}&
{\footnotesize -0.047$\pm0.073$$_{-0.007}^{+0.040}$}&
{\footnotesize 0.099$\pm0.008$$_{-0.032}^{+0.028}$}&
{\footnotesize -}\tabularnewline
\hline 
{\footnotesize 3}&
{\footnotesize 0.8}&
{\footnotesize -0.144$\pm0.057{}_{-0.127}^{+0.217}$}&
{\footnotesize -0.057$\pm0.025$$_{-0.028}^{+0.097}$}&
{\footnotesize 0.124$\pm0.058$$_{-0.137}^{+0.080}$}&
{\footnotesize -0.014$\pm0.032$$_{-0.026}^{+0.051}$}\tabularnewline
\hline 
{\footnotesize 4}&
{\footnotesize 0.6}&
{\footnotesize -0.160$\pm0.027{}_{-0.106}^{+0.111}$}&
{\footnotesize -0.064$\pm0.012{}_{-0.047}^{+0.049}$}&
{\footnotesize 0.143$\pm0.021$$_{-0.057}^{+0.054}$}&
{\footnotesize -0.026$\pm0.008$$_{-0.016}^{+0.017}$}\tabularnewline
\hline
\end{tabular}

~

All the fit results are very consistent with each others. In fits
4 and 6, the smallness of $\mu_{8}$ tends to indicate the convergence
of the OPE series. There is good agreement between the $\Gamma_{1}^{n}$
and $\Gamma_{1}^{p-n}$ analyses and our analysis, although the central
values differ noticeably. Also, our results show the same trend as
the results from the neutron {[}\ref{Meziani}{]} and Bjorken sum
analysis {[}\ref{Bjorken-HT}{]}: The $f_{2}$ coefficient tends to
display an opposite sign as the $\mu_{6}$ coefficient. The alternation
of signs seems to continue with $\mu_{8}$, which indicates that the
overall effects of higher twist are suppressed. This would indicate
that we should expect hadron-parton duality {[}\ref{duality}{]} to
hold for $g_{1}$ at the scale at which the higher twist coefficients
were extracted. The fact that $\mu_{6}$ stands out as the largest
coefficient is, candidly, not surprising since in our $Q^{2}$ ranges,
the $Q^{2}$-behavior is dominated by the elastic contribution which
roughly behaves as $1/Q^{4}$. This feature was also seen in the Bjorken
sum analysis but not in the neutron analysis in which the elastic
contribution is smaller. 

These results can be compared to non-perturbative model predictions:
$f_{2}=-0.037\pm0.006$ {[}\ref{col. pol. 1}{]}, $\mu_{4}/M^{2}=-0.040\pm0.023$
(QCD sum rules {[}\ref{HTSR}{]}), $f_{2}=-0.10\pm0.05$ (MIT bag
model {[}\ref{HTbag}{]}) and $f_{2}=-0.046$ (instanton model {[}\ref{HTinstanton}{]}).
As for the extracted $f_{2}$ and $\mu_{4}$, all the predictions
are negative. The MIT bag model and QCD sum rules agree best with
the fit results, although the other predictions are not ruled out. 

Although agreeing well within uncertainties, $\Gamma_{1}^{p}\neq\Gamma_{1}^{p-n}+\Gamma_{1}^{n}$
from the analyses {[}\ref{Bjorken-HT}{]} and {[}\ref{Meziani}{]}.
This comes from the fact that the $\Delta\Sigma$ extracted from the
proton and neutron analysis are very different: $\Delta\Sigma^{p(n)}=0.15(0.35)$.
This implies that the asymptotic values for the $\Gamma_{1}^{p}$
, $\Gamma_{1}^{n}$ and $\Gamma_{1}^{p-n}$ that anchor the OPE evolutions
used in the fit are inconsistent. As an example, an offset of the
Bjorken sum asymptotic value of $\left(\Delta\Sigma^{p}-\Delta\Sigma^{n}=-0.2\right)/9$
changes the value of $f_{2}^{p-n}$ at $Q^{2}=1$ GeV$^{2}$ by a
factor 2 and the value of $\mu_{6}^{p-n}$ by 50\%.

From the result of fit 1, we can extract the proton color polarizabilities
which are the responses of the color magnetic and electric fields
to the spin of the proton {[}\ref{col. pol. 1},\ref{col. pol. 2}{]}:
$\chi_{E}^{p}=-0.08\pm0.02$ $_{-0.08}^{+0.07}$ and $\chi_{B}^{p}=0.06\pm0.08$$_{-0.04}^{+0.05}$.
As for the neutron, these are of opposite sign and compatible with
zero.

The fact that the higher twist effects are small (at $Q^{2}=1$ GeV$^{2}$)
is somewhat surprising and exciting. However, it implies that accurate
measurements are more delicate: the size of the uncertainty is presently
of the size of the central value itself. In particular, the high energy
missing part, very substantial for the JLab data, introduces a significant
uncertainty. The 12 GeV upgrade of JLab will improve on this issue
and push the $Q^{2}$-coverage of the measurements. It will help in
measuring the higher twist coefficients precisely.

\paragraph{Acknowledgments}

This work was supported by the U.S. Department of Energy (DOE) and
the U.S. National Science Foundation. The Southeastern Universities
Research Association operates the Thomas Jefferson National Accelerator
Facility for the DOE under contract DE-AC05-84ER40150.

\end{document}